\documentclass[aip, amsmath, amssymb, reprint]{revtex4-1}

\usepackage{graphicx}
\usepackage{dcolumn}
\usepackage[utf8]{inputenc}
\usepackage[T1]{fontenc}
\usepackage{mathptmx}
\usepackage{etoolbox}
\usepackage{braket}
\usepackage{accents}
\newcommand*{\bigdott}[1]{%
	\accentset{\mbox{\small\bfseries .}}{#1}}
\usepackage{amsmath}
\usepackage{bm}         
\usepackage{ifpdf}
\usepackage{amssymb,lineno,amsfonts}
\usepackage{bbm}
\usepackage{mathrsfs}
\usepackage{upgreek}
\usepackage{epstopdf}
\usepackage{setspace}
\usepackage{hyperref}
\usepackage[matrix,frame,arrow]{xypic}
\usepackage{float}
\usepackage{soul}
\usepackage{tikz}
\usepackage{color} 

\usepackage{subfigure}
\usepackage{dsfont}
\definecolor{med-blue}{RGB}{25,25,112} 
\hypersetup{colorlinks, linkcolor={blue},citecolor={blue}, urlcolor={black}}

\newcommand{\inpr}[2]{\langle{#1}\vert{#2}\rangle}

\newcommand{\mel}[3]{\langle{#1}\vert{#2}\vert{#3}\rangle}

\newcommand{\tr}{\mathrm{Tr}}

\newcommand{\shr}[0]{Schr\"{o}dinger}
\newcommand{\sheq}[0]{Schr\"{o}dinger equation}

\makeatletter
\def\@email#1#2{%
 \endgroup
 \patchcmd{\titleblock@produce}
  {\frontmatter@RRAPformat}
  {\frontmatter@RRAPformat{\produce@RRAP{*#1\href{mailto:#2}{#2}}}\frontmatter@RRAPformat}
  {}{}
}%
\makeatother
\begin{document}

\preprint{AIP/123-QED}

\title[]{Counterdiabatic driving for long-lived singlet state preparation}

\author{Abhinav Suresh}

\email{abhinav.suresh@students.iiserpune.ac.in, vishal.varma@students.iiserpune.ac.in,}

\email{priya.batra@students.iiserpune.ac.in, mahesh.ts@iiserpune.ac.in}

\author{Vishal Varma}%

\author{Priya Batra}

\author{T S Mahesh}
\affiliation{Department of Physics and NMR Research Center,\\
		Indian Institute of Science Education and Research, Pune 411008, India
}

\date{\today}

\begin{abstract}
The quantum adiabatic method, which maintains populations in their instantaneous eigenstates throughout the state evolution, is an established and often preferred choice for state preparation and manipulation. Though it minimizes the driving cost significantly, its slow speed is a severe limitation in noisy intermediate-scale quantum (NISQ) era technologies. 
Since adiabatic paths are extensive in many physical processes, it is of broader interest to achieve adiabaticity at a much faster rate.  
Shortcuts to adiabaticity techniques which overcome the slow adiabatic process by driving the system faster through non-adiabatic paths have seen increased attention recently.  
The extraordinarily long lifetime of the long-lived singlet states (LLS) in nuclear magnetic resonance, established over the past decade, has opened several important applications ranging from spectroscopy to biomedical imaging. Various methods, including adiabatic methods, are already being used to prepare LLS. In this article, we report the use of counterdiabatic driving (CD) to speed up LLS preparation with faster drives. Using NMR experiments, we show that CD can give stronger LLS order in shorter durations than conventional adiabatic driving.
\end{abstract}

\maketitle

\section{\label{sec:level1}Introduction:\protect\\ }

Adiabatic driving (AD) is the process used to control the dynamics of a quantum system by tuning parameters of the driving Hamiltonian slowly enough that the system continuously remains in its instantaneous eigenstate  \cite{born1928beweis}. This means the system is driven through a sequence of equilibrium states without generating excitations or heat. Such processes are ubiquitous in physics and find a wide range of applications in internal population control\cite{2003JPCA..107.9937D}, state preparation \cite{Rodin_2019}, ion transport in ion traps \cite{article1}, quantum computation \cite{farhi2000quantum}, quantum simulations \cite{Babbush_2014}, in the study of quantum Hall effect\cite{cmp/1104159167} etc. 

\begin{figure}
    \centering
    \includegraphics[trim=3cm 6.2cm 4cm 2cm,width=8.5cm,angle=0,clip=]{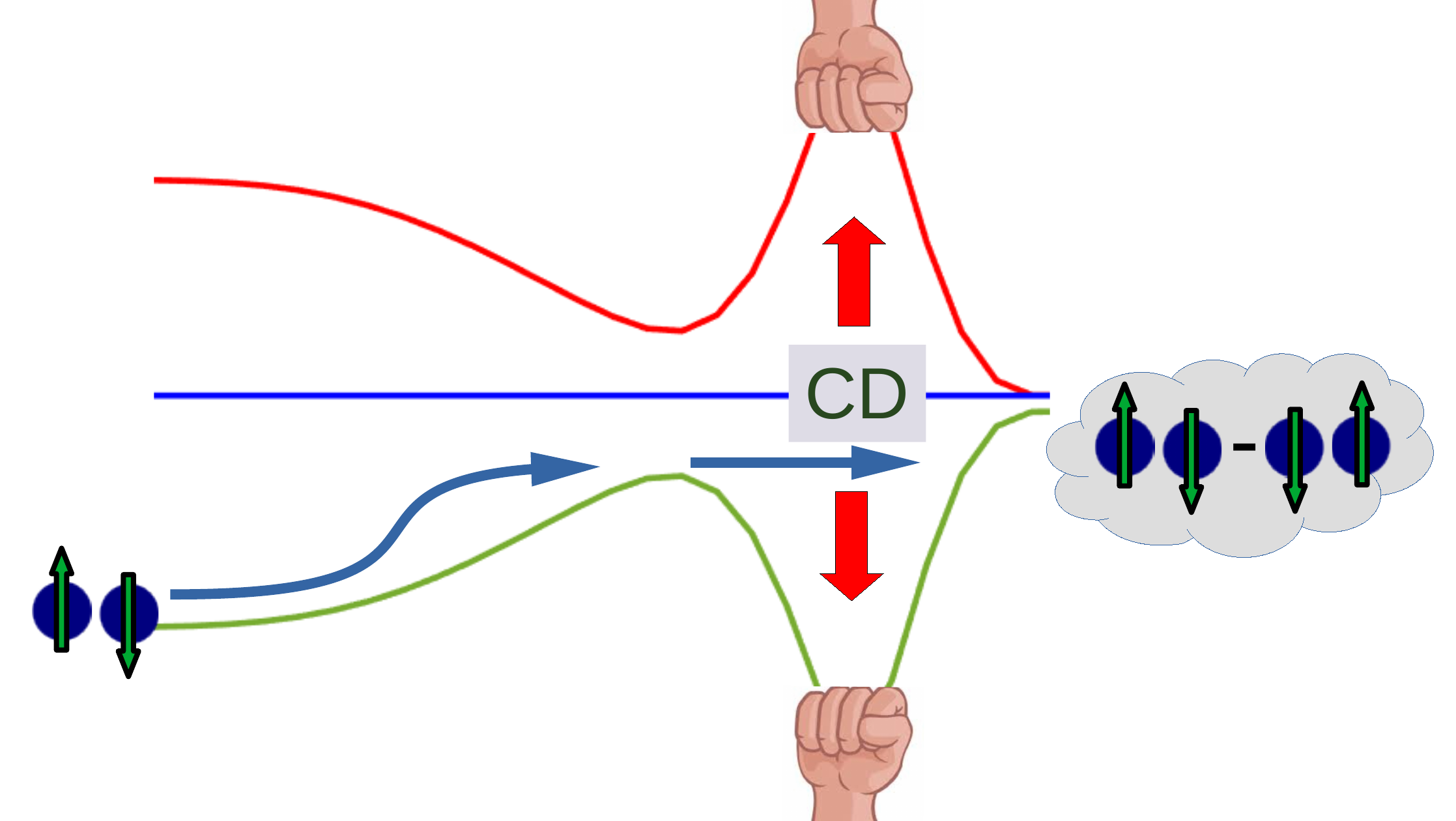}
    \caption{CD pulls the energy levels apart and widens the gap 
    that could be adiabatically traversed faster.
    \label{fig:centralidea}}
\end{figure}

In order to ensure that the system remains in the instantaneous ground state, AD should be as slow as possible \cite{De_Grandi_2010}. This could be a significant disadvantage for noisy intermediate-scale quantum (NISQ) devices which are sensitive to the environment and are much more prone to decoherence. To overcome this limitation of AD, several shortcuts to adiabaticity (STA)\cite{PhysRevLett.104.063002,RevModPhys.91.045001,PhysRevLett.105.123003} techniques have been proposed. STA helps in driving a system from an initial state to the desired state at a faster rate yet preserving the overall adiabaticity. Counterdiabatic driving (CD) or transitionless-quantum driving \cite{Berry_2009} is one such STA method that allows us to engineer the suppression of diabatic transitions caused by faster driving. The central idea of CD is to find an auxiliary Hamiltonian that suppresses the undesirable diabatic transitions by suitably modifying the energy gaps as the system evolves \cite{Campo2013ShortcutsTA} (see Fig. \ref{fig:centralidea}). CD finds its importance in studying quantum dynamics at a timescale smaller than the decoherence time and has been studied in various systems, including quantum heat engines \cite{beau2016scaling,PhysRevE.98.032121}, quantum refrigerators \cite{Funo_2019}, non-equilibrium thermodynamics \cite{PhysRevE.99.022110,PhysRevE.99.032108}, circuit QED \cite{Yin_2022}, and even in evolutionary biology\cite{Iram_2020}. It has been advantageous in quantum annealing  \cite{PhysRevA.95.012309,PhysRevResearch.2.013283}, universal quantum computing \cite{Santos2015SuperadiabaticCE}, population transfer\cite{doi:10.1021/jp030708a}, studying quantum many-body dynamics  \cite{PhysRevA.90.060301,PhysRevLett.109.115703,Hartmann_2019}, in detection of quantum phase transitions \cite{PhysRevA.103.012220} and in quantum approximate algorithms \cite{PhysRevX.11.031070,Wurtz2022counterdiabaticity,PhysRevResearch.4.013141,PhysRevA.105.042415,PhysRevResearch.4.043204}.


Long-lived singlet states (LLS) are of extreme interest to physicists as well as spectroscopists because of their minimal vulnerability to environmental noise. 
While LLS has been discovered \cite{carravetta2004beyond} and extensively studied \cite{pileio2020long} in nuclear magnetic resonance (NMR), similar states have also been explored in other architectures \cite{PhysRevB.95.224105,doi:10.1126/sciadv.aax4482}.
The long-livedness of LLS arises because the antisymmetric singlet state can not couple with symmetric triplet states via symmetry-preserving noise processes such as spin-lattice relaxation mediated by intramolecular dipole-dipole interaction \cite{doi:10.1021/ja0490931}. LLS states are important in a number of scientific areas, including quantum computing, quantum cryptography, and spectroscopy. 
Efficient methods for the creation and manipulation of LLS have applications in medical imaging \cite{theis2016direct,bae201815n4,gloggler2017versatile,mamone2018nuclear}, MRI contrast technique\cite{https://doi.org/10.1002/anie.202014933}, quantum computation \cite{PhysRevA.82.052302}, biomolecular NMR \cite{salvi2012boosting,buratto2014drug,buratto2014exploring,buratto2016ligand} etc. 

The generation of non-equilibrium nuclear singlet states was first demonstrated in weakly coupled molecules by Carravetta and Levitt\cite{doi:10.1021/ja0490931}. Since then, several methods \cite{pileio2020long}, including an adiabatic method \cite{Rodin_2019} for LLS preparation, have been proposed and demonstrated. In this article, we report the use of CD for faster LLS generation.  

In Sec. II, we first explain the theoretical framework of adiabatic driving (AD), followed by a brief theory of CD. In Sec. III, we apply CD on a two-qubit system, provide numerical and experimental simulations of the CD pulse sequence, and discuss their results. Finally, we summarize and conclude in Sec. IV.

\section{Theory}
\subsection{Adiabatic driving (AD)}
Let us consider a time-dependent Hamiltonian $H(t)$ with instantaneous eigenvalues $\{E_n(t)\}$ and eigenstates $\{\ket{n(t)}\}$.  A general instantaneous state $\ket{\Psi(t)}$ can be expanded in the complete eigenbasis, in $\hbar=1$ units, as
\begin{align}
\ket{\Psi(t)} =  \sum_n c_n(t) e^{i\theta_n(t)} \ket{n(t)},
~\mbox{where}~
\theta_n(t) = -
\int_0^{t} dt' E_n(t').
\end{align}
We now try to follow the dynamics of coefficients $c_n(t)$.
Substituting the above in \sheq
\begin{align}
i \frac{\partial}{\partial t}
\ket{\Psi(t)} = H(t) \ket{\Psi(t)},
\end{align}
we obtain
\begin{align}
i \sum_n 
\left[
\dot{c_n} \ket{n} 
+ {c_n} \ket{\dot{n}}
+i\dot{\theta}_n {c_n} \ket{n} 
\right] e^{i\theta_n}
&=\sum_n E_n c_n \ket{n} e^{i\theta_n},
\nonumber \\
\mbox{or,}~~ \sum_n 
\left[
\dot{c_n} \ket{n} 
+ {c_n} \ket{\dot{n}}
\right]e^{i\theta_n}
&=0.
\nonumber \\
\therefore
~~  \sum_n 
\left[
\dot{c_n} \inpr{m}{n} 
+ {c_n} \inpr{m}{\dot{n}}
\right]e^{i\theta_n}
&=0,
\nonumber \\
\mbox{i.e.,}~
 \dot{c_m} 
= - {c_m} \inpr{m}{\dot{m}}-\sum_{n\neq m} {c_n} \inpr{m}{\dot{n}}
e^{i(\theta_n-\theta_m)}
.
\label{eq:ad1}
\end{align}
To determine $\inpr{m}{\dot{n}}$ for nondegenerate levels $m \neq n$, 
we differentiate the \shr{} equation $H\ket{n} = E_n\ket{n}$ and take inner product with $\ket{m}$,
\begin{align}
\mel{m}{\dot{H}}{n} + E_m\inpr{m}{\dot{n}} &= \dot{E}_m \inpr{m}{n}
+ E_n \inpr{m}{\dot{n}},
\nonumber \\
\mbox{or}~~
\inpr{m}{\dot{n}} &= \frac{\mel{m}{\dot{H}}{n}}{E_n-E_m}.
\label{transition}
\end{align}

The adiabatic approximation involves setting  \cite{griffiths2018introduction}
\begin{align}
&\dfrac{|\mel{m(t)}{\dot{H}(t)}{n(t)}|}{|E_n(t)-E_m(t)|}  \approx 0 ~~\mbox{for all $t$, s.t. Eq. \ref{eq:ad1} yields}
\label{eq:adcondn}
\\
& \dot{c_m} = -c_m \inpr{m}{\dot{m}},~~\mbox{with solution}~
c_m(t) = c_m(0) e^{i\gamma_m(t)},~
\nonumber \\
&\mbox{where}~
\gamma_m(t) \approx i\int_0^t dt' \langle m(t') 
\vert \dot{m}(t') \rangle.
\label{eq:ad3}
\end{align}
In the above, $\theta_n$ and $\gamma_n$ are the dynamical and geometric phases, respectively.
If the system starts from $\ket{n(0)}$, such that $c_n(0)=1$ and $c_m(0)=0$ for all $m\neq n$, then we obtain
\begin{equation}
\ket{\Psi(t)} = e^{i\theta_n(t)}e^{i\gamma_n(t)}\ket{n(t)}.
\label{eq:adstate}
\end{equation}

Our goal is an adiabatic evolution of a system that starts from an eigenstate of a given initial Hamiltonian $H_I$ at $t=0$ and reaches the corresponding eigenstate of the final Hamiltonian $H_F$ at time $t = T$. The instantaneous adiabatic Hamiltonian is parameterized as
\begin{align}
H_{AD}(t) = (1-\lambda(t)) H_I + \lambda(t) H_F,
\label{eq:Had}
\end{align}
where $\lambda(0) = 0$ and $\lambda(T) = 1$.  This way, one can tune the Hamiltonian from the initial form $H_I$ to the final form $H_F$ via the scalar parameter $\lambda$. However, this must be done slowly to satisfy adiabaticity. In the following, we discuss the CD method, which overcomes this limitation.

\subsection{Counterdiabatic driving (CD)}
Although adiabatic control has its merits, the adiabatic approximation limits the maximum speed of the drive.
If the drive Hamiltonian $H_{AD}(t)$ varies too fast compared to the energy difference so that the adiabatic approximation of Eq. \ref{eq:adcondn} is not satisfied, then undesired diabatic transitions may occur.
CD helps in the suppression of diabatic transitions due to the fast driving, and thereby, it allows for shorter control sequences.
Let $U_\lambda$ be the change of basis operator to the moving frame generated by $H_{AD}$. The transformation to the moving frame (denoted by $\sim$) and corresponding \shr{} equation can be written as \cite{geometrycd}
\begin{align}
\ket{\Psi(t)}&= U_\lambda\ket{\widetilde\Psi(t)} ~\mbox{and}
\nonumber\\
i \frac{\partial}{\partial t}U_\lambda \ket{\widetilde\Psi(t)}
&=H_{AD} U_\lambda \ket{\widetilde\Psi(t)},
\nonumber\\
\mbox{i.e.,}~~ i \left[\dot{\lambda} \frac{\partial U_\lambda}{\partial \lambda} \ket{\widetilde\Psi}+ U_\lambda \frac{\partial \ket{\widetilde\Psi}}{\partial t}\right]&=H_{AD} U_\lambda \ket{\widetilde\Psi(t)}.
\end{align}

Multiplying both sides by $U_\lambda^\dagger$ and rearranging terms, we get \shr{} equation in the moving frame
\begin{align}
i \frac{\partial \ket{\widetilde\Psi(t)}}{\partial t}&=\widetilde{H}_{AD}\ket{\widetilde{\Psi}(t)},
\nonumber \\
\mbox{where}~~
\widetilde{H}_{AD}&= U_\lambda^{\dagger}H_{AD}U_\lambda
-i\dot{\lambda}U_\lambda^{\dagger}\frac{\partial U_\lambda}{\partial \lambda}.
\end{align}
Since $U_\lambda$ is generated by $H_{AD}$,
 the effective Hamiltonian in the moving frame is of the form
 \begin{align}
\widetilde{H}_{AD}(t)=H_{AD}(t) -\dot{\lambda}A_\lambda(t),~~
\mbox{where}~~ A_{\lambda}=i U_\lambda^\dagger\frac{\partial U_\lambda}{\partial \lambda}
 \end{align}
is the adiabatic gauge potential. In general, a gauge potential generates translations in the parameter space.
The second term $-\dot{\lambda}A_\lambda$ arises from the non-inertial frame transformation and is responsible for the diabatic evolution. This could be seen by taking the matrix element of $A_\lambda$ in the lab frame \cite{geometrycd}, namely
\begin{align}
\inpr{m}{A_\lambda| n}=i\frac{\inpr{m}{\partial_\lambda H |n}}{E_n-E_m}
\end{align}
which is nothing but $\inpr{m}{\dot{n}}$ parameterized by $\lambda$ in Eq. \ref{transition}.
To counter these diabatic transitions we add the same term $\dot{\lambda}A_\lambda(t)$ to the lab-frame control Hamiltonian $H_{AD}$ to form total Hamiltonian  \cite{geometrycd}
\begin{equation}
    {H_{CD}(t)}=H_{AD}(t)+H_\lambda(t),~~\mbox{where}~~
    H_\lambda(t) = \dot{\lambda}(t)A_\lambda(t).
    \label{eq:hcd}
\end{equation}
Here $H_\lambda(t)$ is the correction term which ensures that the eigenstates of the total Hamiltonian $H_{CD}$ are of the form given in Eq. \ref{eq:adstate} because it renders the total moving frame Hamiltonian to be diagonal and inertial. This way, the diabatic transitions are suppressed in CD by finding the appropriate adiabatic gauge potential.

In general, setting up the exact gauge potential requires complete spectral knowledge at every instant of time.  It satisfies the condition \cite{geometrycd}
\begin{align}
\left[H_{AD},M_\lambda \right]=0, ~~\mbox{where} ~~
M_\lambda = \frac{\partial H_{AD}}{\partial \lambda}-i[H_{AD},A_\lambda].
\end{align}
But, since the form of $\lambda$ is not known apriori, it is essentially an optimization problem.  For a chosen ansatz of $\lambda$, an approximate form of the adiabatic gauge potential up to first-order variational coefficient
$\alpha$ is \cite{Claeys2019FloquetEngineeringCP}
\begin{align}
A^{^{(1)}}_\lambda & =  i\alpha\left[H_{AD},\frac{\partial H_{AD}}{\partial\lambda} \right]  
\nonumber \\
&=  
i\alpha\left[H_{AD},(H_F-H_I) \right]
=i\alpha 
\left[H_I,H_F\right], ~~\mbox{such that}
\nonumber \\
[H_{AD},G_\lambda] &\approx 0, ~~\mbox{where}~~
G_\lambda=\frac{\partial H_{AD}}{\partial \lambda}-i[H_{AD},A^{^{(1)}}_\lambda].
\label{eq:alambda}
\end{align}
In order to find the optimal $A_\lambda^{(1)}$, we find an appropriate variational coefficient $\alpha$ that minimizes the trace distance between $M_\lambda$ and $G_\lambda$.
This minimization was shown to be equivalent to 
minimizing the action \cite{geometrycd}
\begin{align}
S & =\tr\left(G_\lambda^2\right)
= \tr\left((
H_F-H_I + \alpha
\left[
H_{AD},
\left[
H_I,H_F
\right]
\right]
\right)^2).
\label{eq:seq}
\end{align}                                                                 
In the following, we describe implementing the above protocol on a two-qubit system.

\subsection{AD and CD on a two-qubit system}
We now consider a two-qubit system with Hamiltonian,
\begin{align}
  H_I=-\pi\delta \left(I_{1z}-I_{2z}\right)+2\pi J
  \mathbf{I}_{1} \cdot \mathbf{I}_{2},
  \label{eq:twoqham}
\end{align}
where $\mathbf{I}_i$ are the vector spin operators with components $I_{i\beta}$ with $i\in\{1,2\}$ and $\beta \in \{x,y,z\}$.

We first set up $H_{AD}$ in the form of Eq. \ref{eq:Had} with the final Hamiltonian 
\begin{align}
H(T) = H_F = 2\pi J  \mathbf{I}_{1} \cdot \mathbf{I}_{2}.
\end{align}

The control parameter $\lambda$ should obey the STA conditions \cite{PhysRevA.83.013415},
\begin{align}
\lambda(0)=0, ~ \lambda(T)=1,~\dot{\lambda}(0)=\dot{\lambda}(T)=
\ddot{\lambda}(0)= \ddot{\lambda}(T)=0.
\end{align}
An anzast that satisfies these conditions is \cite{_epait__2023}  \begin{equation}
\lambda(t)={\sin^2 \left(\frac{\pi}{2}\sin^2 \frac{\pi \,t}{2\,T}\right)}.
\label{eq:lambdaform}
\end{equation}

\begin{figure} 
    \includegraphics[width=8cm]{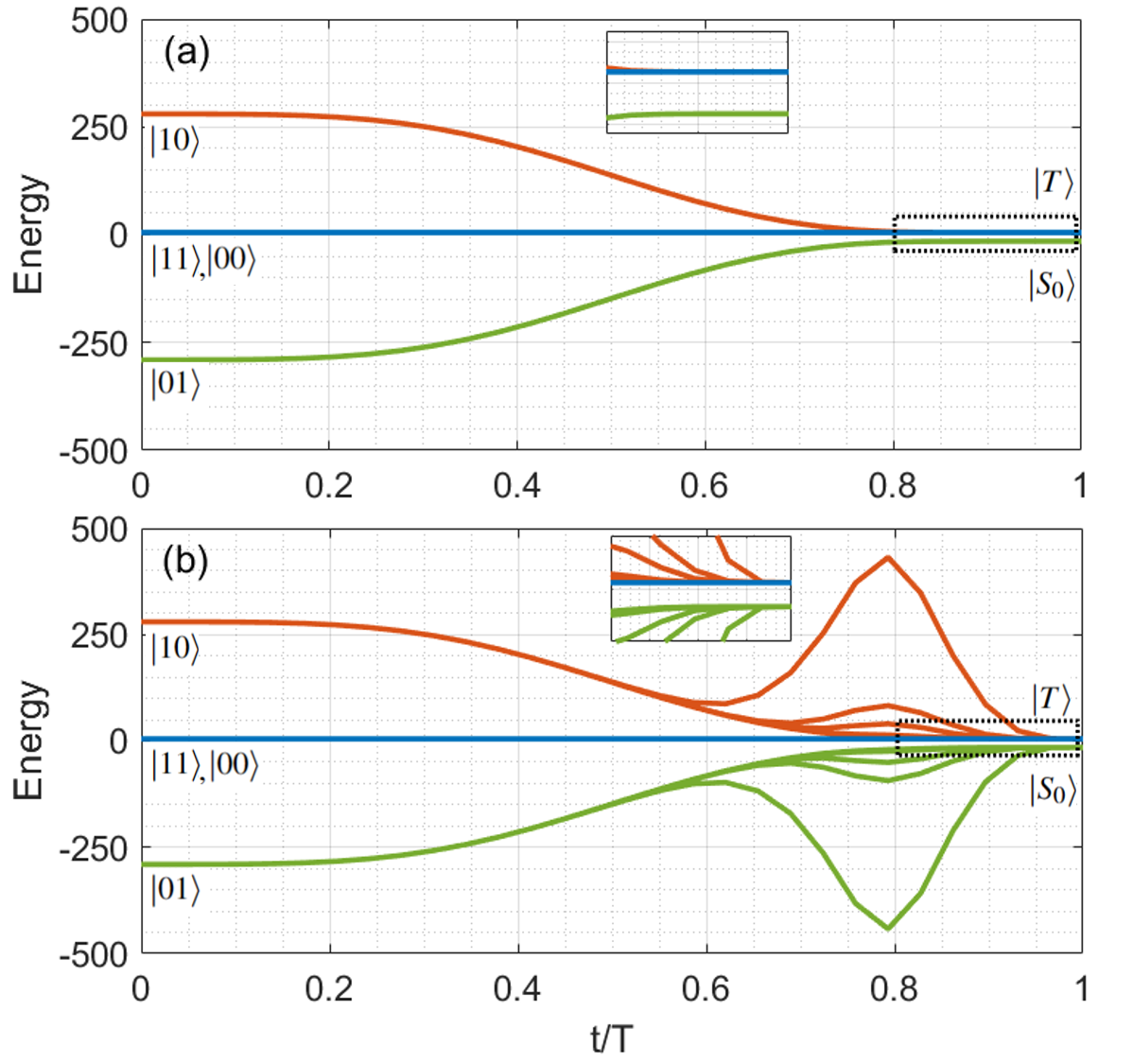}
    \caption{Evolution of eigenvalues for (a) AD (b) CD.  In (b), five sets of eigenvalues correspond to the total time  $T=\{0.01,0.05,0.1,0.3,0.5\}$, $\delta = 90.7$ Hz, and $J=3.24$ Hz. The furthest stretched-out eigenvalues correspond to the shortest $T$. AD eigenvalues have no visible change w.r.t. $T$. The inset shows a zoomed-in portion from 0.8 to 1 t/T where the energy gap is visible.}
    \label{fig:eigval}
\end{figure}

Fig. \ref{fig:eigval} (a) represents the evolution of all the four eigenvalues of $H_{AD}(t)$ (see Eq. \ref{eq:Had}) for the above forms of $H_I$, $H_F$, and $\lambda(t)$. Here the initial ground state is $\ket{01}$ in the computational basis.  As desired, the ground state of the final Hamiltonian $H_F$ is the singlet state,
\begin{align}
\ket{S_0} = \frac{\ket{01}-\ket{10}}{\sqrt{2}}.
\end{align}
Note that the ground state always remains nondegenerate, even for the final Hamiltonian $H_F$ (see inset of Fig. \ref{fig:eigval} (a)). However, it is clear that the energy gap between the ground state and the excited states decreases asymptotically, thereby limiting the speed of the AD drive. 

\begin{figure}
    \includegraphics[width=8cm]{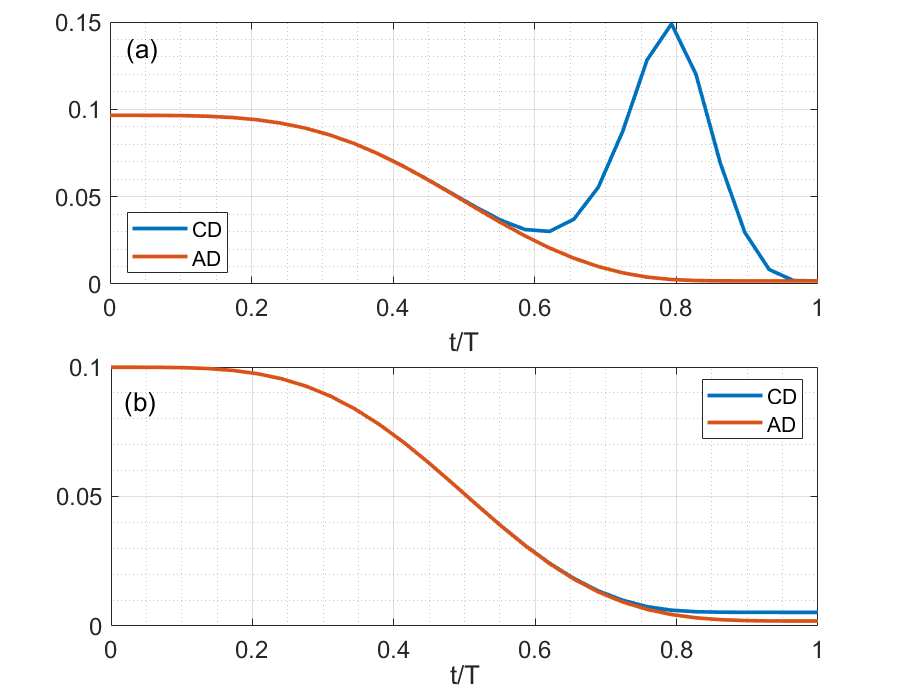}
    \caption{Evolution of (a) the dynamical phase and (b) the geometric phase for simulation parameters $T=0.01$, $N=30$, $\delta = 90.7$ Hz, and $J=3.24$ Hz. }
    \label{fig:dyngeophase}
\end{figure}

Now we shall construct the CD drive.
For the above system and the form of $\lambda(t)$ in Eq. \ref{eq:lambdaform}, the first-order adiabatic gauge potential from Eq. \ref{eq:alambda} becomes
\begin{align}
A_\lambda(t)&=i\alpha(t)[H_I,H_F]
\nonumber \\
&= -4\pi^2(\delta ) (J)\alpha(t) (I_{1x} I_{2y}-I_{1y}I_{2x}).
\label{eq:alambda1}
 \end{align}
 On minimizing the action $S$ via Eq. \ref{eq:seq} we obtain the first-order variational coefficient $\alpha$ as
  \begin{align}
 \alpha(t)=\frac{1}{4\pi^2\{4\delta^2(\lambda(t)-1)^2+J^2\}}.
 \label{eq:alpha}
 \end{align}
 Now, using Eqs. \ref{eq:alambda1}  and \ref{eq:alpha} in Eq. \ref{eq:hcd}, we get the auxiliary term $H_\lambda$ as 
 \begin{align}
 H_\lambda(t) 
  &= \kappa(t) (I_{1x} I_{2y}-I_{1y}I_{2x}), ~~
  \mbox{where}
  \nonumber \\
  \kappa(t) &=  
 \frac{-\bigdott{\lambda}(t) \delta J}{4\delta^2(\lambda(t)-1)^2+J^2}
 \label{eq:Hlambdakappa}
 \end{align}
 is the amplitude of the $H_\lambda$ drive.
 By adding this to $H_{AD}$, we obtain $H_{CD}$ that would enable faster driving without diabatic transitions. To understand this improvement, we look at the evolution of all the eigenvalues of $H_{CD}(t)$ for different total times $T$ represented in Fig. \ref{fig:eigval} (b). It is evident that, compared to the AD case, the energy levels are pulled apart, facilitating faster driving. In fact, the effect is more dramatic as $T$ becomes shorter. This behaviour explains why CD can be advantageous for faster state preparations.

Fig. \ref{fig:dyngeophase} shows the evolution of dynamical phase $\theta_{\ket{01}}(t)$ and geometric phase $\gamma_{\ket{01}}(t)$ for the two-qubit system.  As expected, the dynamical phase shows a strong peak coinciding with the widening of the energy gap in Fig. \ref{fig:eigval}.  Interestingly, the geometric phase has no such strong feature.

\section{Simulations and Experiments}
\subsection{Two-qubit NMR system}
\begin{figure}
\includegraphics[trim=4cm 0cm 3cm 1.3cm,width=8cm,clip=]{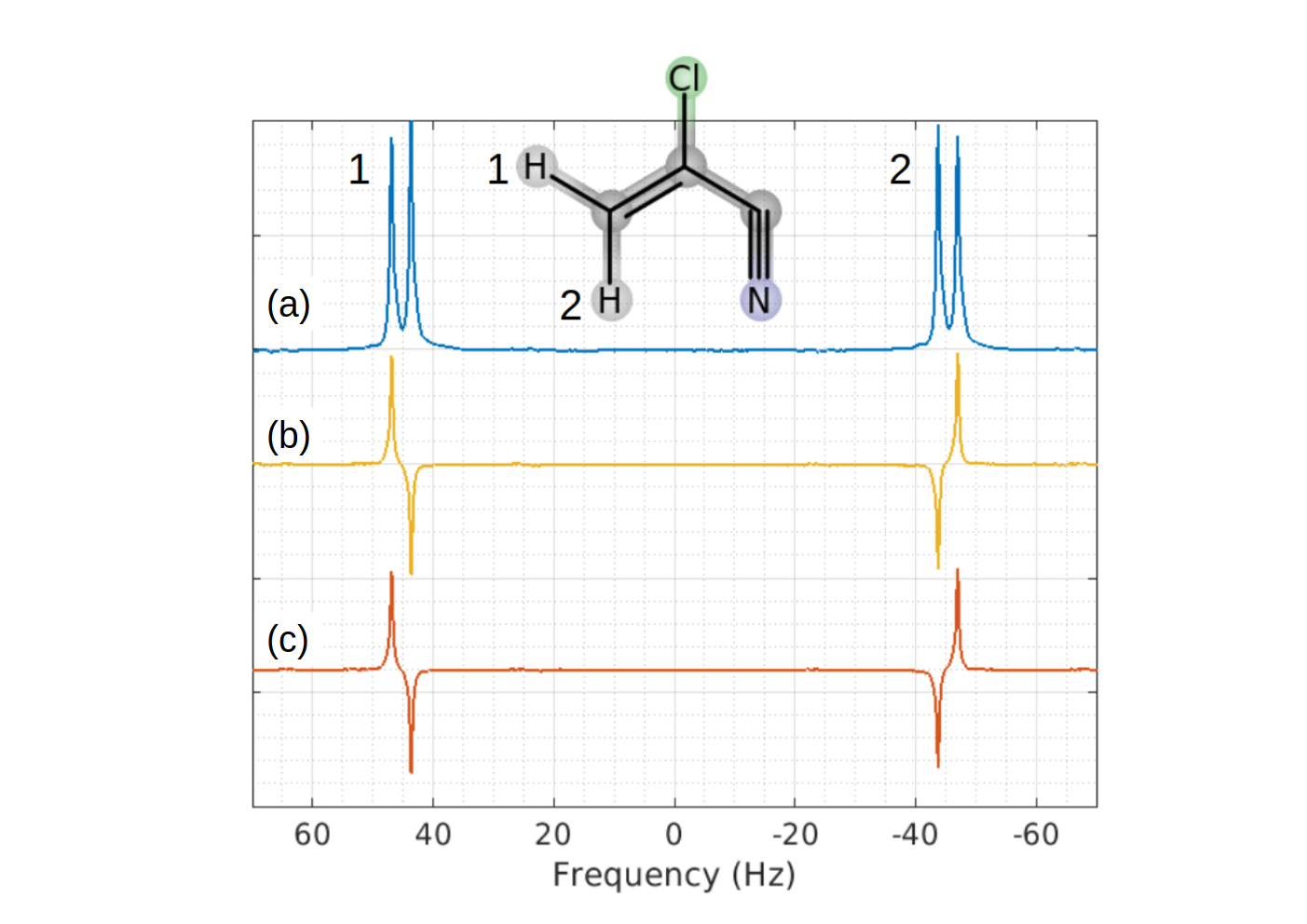}
\caption{\label{fig:system} (a) Spectra of 2-Chloroacrylonitrile (CAN), (b) LLS spectrum with AD ($T=0.1$ s, $N=15$, $\tau_S = 30$ s), (c) LLS spectrum with CD ($T=0.5$ s, $N=30$, $\tau_S = 30$ s)}
\end{figure}
Our two-qubit sample consists of $20\mu l$ of 2-chloroacrylonitrile (CAN) dissolved in $600 \mu l$ deuterated dimethylsulphoxide (DMSO). We treat the two spin-1/2 protons of CAN as two qubits (see inset of Fig. \ref{fig:system} (a)). 
The experiments were carried out on a Bruker 500 MHz NMR spectrometer at an ambient temperature of 300 K. 
The rotating frame Hamiltonian of the two-qubit system is the same form as Eq. \ref{eq:twoqham},
with $\delta = 90.7$ Hz being the chemical shift difference and $J = 3.24$ Hz being the strength of the indirect spin-spin interaction. The one-pulse $^1$H spectrum of CAN is shown in Fig. \ref{fig:system} (a).

\subsection{The NMR pulse-sequence}
\begin{figure}
\includegraphics[trim=5cm 1.5cm 5cm 2.5cm,width=9cm,angle=0,clip=]{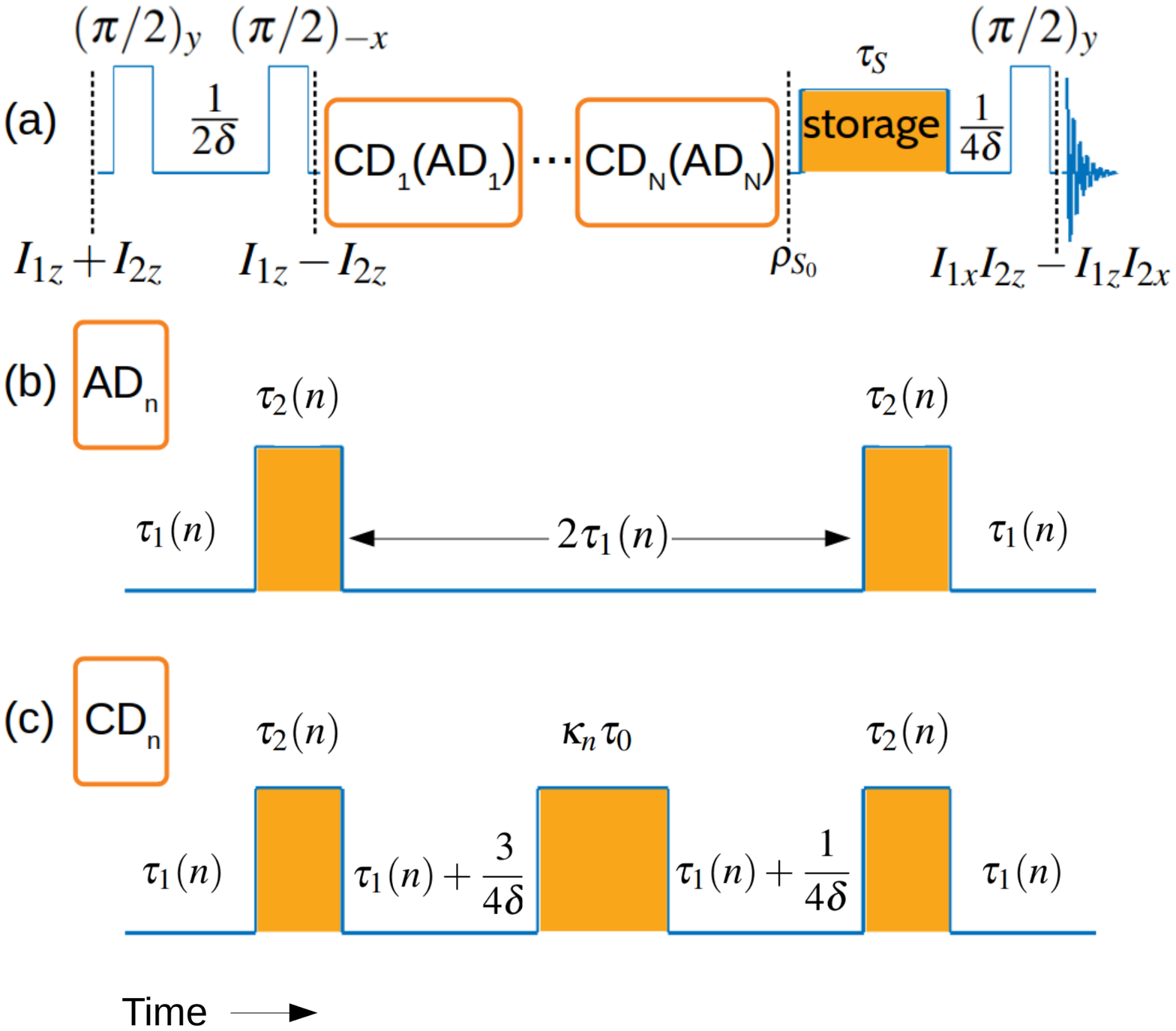}
\caption{\label{fig:pulseq}a) Total pulse sequence, b) AD Pulse sequence, c) CD pulse sequence,
where $\tau_S$ is the storage time. The filled rectangles correspond to WALTZ-16 sequences.
}
\end{figure}

The NMR pulse sequence is schematically illustrated in Fig. \ref{fig:pulseq}.  It has the following stages.

\subsubsection{Initialization}
As seen in Fig. \ref{fig:eigval}, the singlet state $\ket{S_0}$ is adiabatically connected to the ground state $\ket{01}$ of the initial Hamiltonian $H_I$. 
In NMR, the thermal initial state and the subsequently prepared non-equilibrium states are highly mixed.
In the mixed states
\begin{align}
\rho_{_{1}} &= I_{1z} - I_{2z}~~
\mbox{and}~~ 
\rho_{S_0} = 
-\mathbf{I}_1\cdot \mathbf{I}_2,
\end{align}
the excess populations are in basis states $\ket{01}$ and the $\ket{S_0}$ respectively.
As shown in Fig. \ref{fig:pulseq} (a), we use a pair of hard pulses separated by a delay to prepare the initial state $\rho_{_{1}}$ starting from the thermal state $\rho_{_{0}} = I_{1z}+I_{2z}$.  
The task now is to start from $\rho_{_{1}}$ and reach $\rho_{S_0}$ using either AD or CD evolutions, which are described below.

\subsubsection{AD and CD evolutions}
We first discretize $\lambda$ and $\kappa$ of Eqs. \ref{eq:lambdaform} and \ref{eq:Hlambdakappa} into $N$ segments each of duration $\tau_0$ such that $N\tau_0 = T$ and
\begin{align}
\lambda_n &=  \sin^2 \left(\frac{\pi}{2}\sin^2 \frac{\pi \,n}{2\,N}\right),~~
\mbox{and}
\nonumber \\
\kappa_n &= \frac{-\dot{\lambda_n}\delta J}{4\delta^2(\lambda_n-1)^2+J^2}.
\label{eq:lambdakappadiscrete}
\end{align}
The corresponding piecewise continuous AD and CD Hamiltonians \begin{align}
H_{AD}^{\lambda_n} & = 
(1-\lambda_n)H_I + \lambda_n H_F~~\mbox{and}
\nonumber \\
H_{CD}^{\lambda_n} & = H_{AD}^{\lambda_n} + H_{\lambda_n},
\label{eq:hadcddiscrete}
\end{align}
which are the discretized forms of Eqs. \ref{eq:Had} and \ref{eq:hcd}.

As illustrated in Fig. \ref{fig:pulseq} (a), the total evolutions under AD and CD can be expanded in terms of $N$ segments as
\begin{align}
U_{AD} &\approx \prod_{n=1}^N U_{AD}^{\lambda_n},~~\mbox{where}~~
U_{AD}^{\lambda_n} = 
\exp\left(
-i H_{AD}^{\lambda_n}\tau_0
\right)
\nonumber \\
\mbox{and}~~U_{CD} &\approx  \prod_{n=1}^N U_{CD}^{\lambda_n}
,~~\mbox{where}~~
U_{CD}^{\lambda_n} = 
\exp\left(
-i H_{CD}^{\lambda_n}\tau_0
\right).
\label{eq:uaducd}
\end{align}

\begin{figure*}
    \centering
    \includegraphics[trim=0cm 2cm 0cm 5cm,width=18cm,clip=]{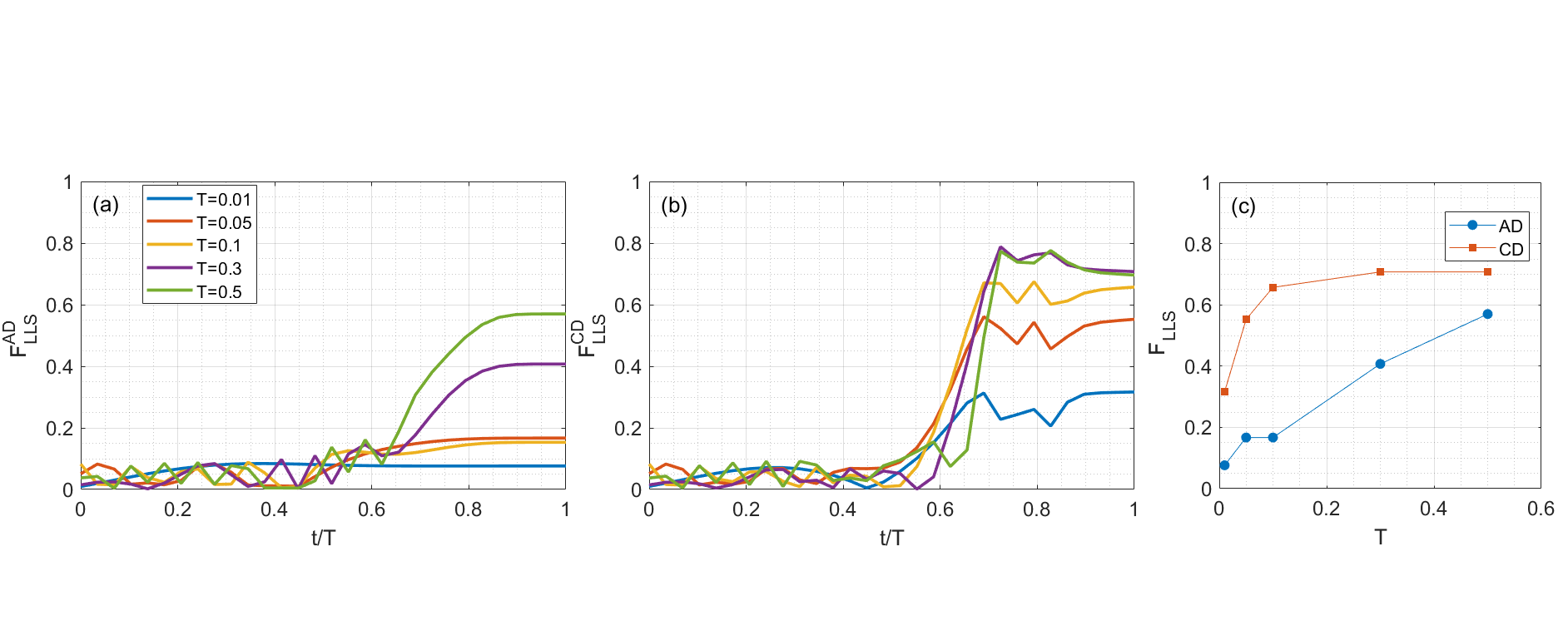}

    \caption{Numerical simulations of AD (a) and CD (b) evolutions by the pulse sequences shown in Fig. \ref{fig:pulseq} (b,c), applied on mixed initial states available in NMR systems. (c) Final fidelity w.r.t. total time $T$ for both AD and CD. The number of iterations was kept at N=30 for all the simulations.}
    \label{fig:fidelity}
\end{figure*}
Now we describe how we can realize the AD and CD segments $U_{AD}^{\lambda_n}$ and $U_{CD}^{\lambda_n}$. As shown in Fig. \ref{fig:pulseq} (b), we stack a pair of AD sequences, between which the auxiliary unitary for CD can be inserted. This allows a straightforward comparison between the two methods. To this end, we set up AD and CD segments in the form
\begin{align}
U_{AD}^{\lambda_n} &= U_{AD}^{\lambda_n/2} 
~ \mathbbm{1} ~
U_{AD}^{\lambda_n/2}~~
\mbox{and}
\nonumber \\
U_{CD}^{\lambda_n} &= U_{AD}^{\lambda_n/2} 
~ W^{\lambda_n} ~
U_{AD}^{\lambda_n/2},
\end{align}
where $\mathbbm{1}$ is the identity operator and $W^{\lambda_n}$ is the propagator for the auxiliary Hamiltonian $H_{\lambda_n}$ in Eq. \ref{eq:hadcddiscrete} implemented via symmetrized Trotter decomposition.
The propagator $U_{AD}^{\lambda_n/2}$ for adiabatic evolution in Eq. \ref{eq:Had} is implemented again via symmetrized Trotter decomposition as follows:
\begin{align}
U_{AD}^{\lambda_n/2} &= 
U_I^{{(n)}} ~
U_F^{{(n)}} ~ 
U_I^{{(n)}},
\nonumber \\
U_I^{{(n)}} &= \exp\left(-iH_I\tau_1{(n)}\right),
\nonumber \\
U_F^{{(n)}} &= \exp\left(-iH_F\tau_2{(n)}\right),
\nonumber \\
 \tau_1{(n)}&=\frac{(1-\lambda_n)\tau_0}{4}, 
~~\mbox{and}~~ \tau_2{(n)}=\frac{\lambda_n\tau_0}{2}.
\label{eq:taus}
\end{align}
As mentioned above, $H_I$ is the natural NMR Hamiltonian for the weakly coupled spin pair, and therefore $U_I^{(n)}$ is implemented simply with a delay $\tau_1(n)$. An effective $H_F$ Hamiltonian is realized by suppressing the chemical shift. In our experiments, we used WALTZ-16 spin lock for duration $\tau_2(n)$ to implement $U_F^{(n)}$.

To implement CD segment, we need to insert $W^{\lambda_n}$ as described in Eq. \ref{eq:hadcddiscrete}.
One way to realize $W^{\lambda_n}$ is with the help of an R.F. rotation $U_{x}^\alpha$ followed by chemical shift  $U_\delta^\beta$ and weakly coupled J-evolution transformations $U_J^\eta$. These transformations are of the form
\begin{align}
U_{x}^\alpha &= \exp\left[-i\alpha (I_{1x}+I_{2x})\right],
\nonumber \\
U_\delta^\beta &= \exp\left[
-i \beta (-I_{1z}+I_{2z})
\right],  ~~\mbox{and} 
\nonumber \\
U_J^\eta &= \exp\left[
-i \eta (2I_{1z}I_{2z})
\right].   
\end{align}
The rotation $\alpha$ is realized by a field of amplitude $\nu_{x} \gg \delta$ and duration $\tau_{x} = \alpha/(2\pi\nu_{x})$ applied on both the spins.
The rotation $\beta$ is realized simply by a delay $\tau_\delta = \beta/(\pi\delta)$, by assuming that $\delta \gg J$.
The rotation $\eta$ is realized by a spin-echo sequence $\tau_J \_ U_x^\pi \_ \tau_J $ with the delay $\tau_J = \eta/(2\pi J)$.  
With these transformations, the auxiliary unitary $W^\lambda$ is given by
\begin{align}
 W^{\lambda_n} = U_{x}^{\pi/2}~
 U_J^{\pi/2} ~
 U_\delta^{\pi/2} ~ U_{x}^{\kappa_n} ~
 U_\delta^{3\pi/2} ~
 U_J^{3\pi/2} ~ 
 U_{-x}^{\pi/2}.
 \label{eq:wlambda1}
\end{align}
For systems with $J \ll \delta$, the physical duration of the above unitary is $\sim 2/J$ for each iterative segment. Considering the finite relaxation times of the system, this sequence may be prohibitively long when incorporated in the iterative CD sequence of Eq. \ref{eq:uaducd}, and therefore we discard this method.

A much shorter but approximate sequence (which is accurate up to an extra $I_{1z}I_{2z}$ term) is
\begin{align}
W^{\lambda_n} \sim U_\delta^{\pi/4}~U_F^{\kappa_n}~U_\delta^{3\pi/4}.  
\label{eq:approx_seq}
\end{align}
Here $U_F^\kappa$ is realised by WALTZ-16 spin lock of amplitude $\nu_F$ for a duration of $\kappa_n/(2\pi \nu_F)$.  For $\nu_F \gg \delta$, this sequence takes a duration of only $1/\delta$, which is much shorter than the one described in Eq. \ref{eq:wlambda1}. This sequence, illustrated in Fig. \ref{fig:pulseq} (c), is used in our experiments.

First, we numerically simulate the ideal unitary evolution from the pure ground state of $H_I$, keeping fidelity 
\begin{align}
F_{LLS}(t) =  
\left\vert
\frac{\tr[\rho(t) ~\rho_{S_0}]}{\sqrt{\tr[\rho^2(t)] ~\tr[\rho^2_{S_0}]}}
\right\vert
\end{align}
of the instantaneous state $\rho(t)$ with LLS $\rho_{S_0}$ as a benchmark for state evolution.  

Here $\rho(t)$ is either
\begin{align}
\rho^{AD}(t) &= \prod_{n=1}^N U_{AD}^{\lambda_n} ~ \rho_1 ~{(U_{AD}^{\lambda_n})}^\dagger,~~\mbox{or}, 
\nonumber \\
\rho^{CD}(t) &= \prod_{n=1}^N U_{CD}^{\lambda_n} ~ \rho_1 ~{(U_{CD}^{\lambda_n})}^\dagger.\end{align}
We numerically simulate the pulse sequences in
Fig. \ref{fig:pulseq} (b,c) for five different values of the total time duration $T$. The results for AD and CD are shown respectively in Figs. \ref{fig:fidelity} (a) and (b).  Fig. \ref{fig:fidelity} (c) plots the final fidelity of LLS for each value of $T$.
It is clear that the fidelity of LLS in CD raises quickly with $T$.  However for long total durations of $T$, the evolution is sufficiently adiabatic, and the superiority of CD over AD fades away. 

\subsubsection{LLS storage and detection}
We now store the LLS for a storage time $\tau_S$ under isotropic Hamiltonian $H_F$ again using a WALTZ-16 spinlock. Finally, we detect LLS after converting it into observable single-quantum magnetization $I_{1x}I_{2z}-I_{1z}I_{2x}$ using a delay and $(\pi/2)$ pulse as shown in Fig. \ref{fig:pulseq}
 (a). This produces a characteristic anti-phase signal as shown in Fig. \ref{fig:system} (b) and (c).
 In the following, we benchmark the performance of our CD pulse sequence (\ref{fig:pulseq} (c)) with the AD sequence (\ref{fig:pulseq} (b)) via liquid-state NMR experiments.

\subsection{Experimental results}
\begin{figure}
    \includegraphics[trim=0cm 0cm 0cm 0cm,width=8cm,clip=,]{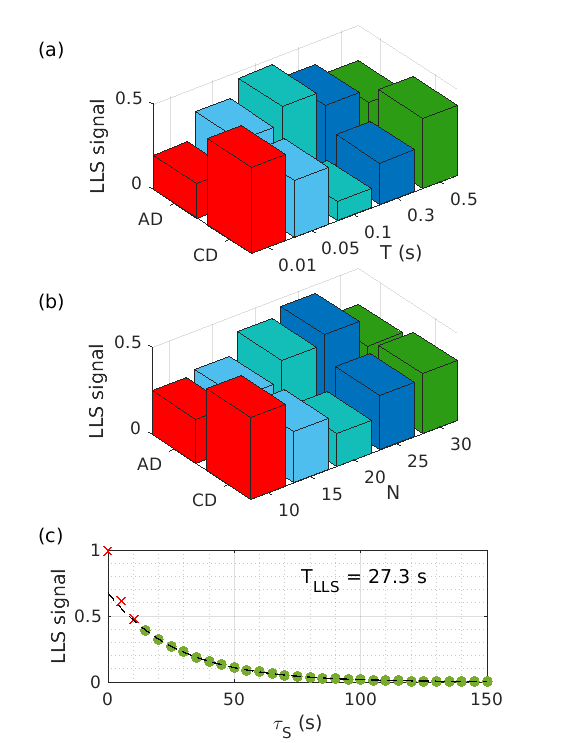}
    \caption{(a,b) Comparison of performances of AD and CD for (a) varying total time $T$, obtained by averaging signal intensities from five different experiments with different iteration number $N$ and (b) varying iteration number $N$  obtained by averaging signal intensities from five different experiments with different total time $T$. As in simulations of Figs. \ref{fig:eigval} and \ref{fig:fidelity}, we have kept $T\in\{0.01,0.05,0.1,0.3,0.5\}$ and $N\in\{10,15,20,25,30\}$. The storage time $\tau_s$ was kept at 30s for all experiments.
    (c) Decay of LLS with the storage time $\tau_s$ showing the LLS lifetime of 27.3 s.
    }
    \label{cd_ad_bar}
\end{figure}

The physical implementation times of AD and CD evolutions are (see Fig. \ref{fig:pulseq} and Eq. \ref{eq:taus})
\begin{align}
T^{AD}(\tau_0) &= \sum_{n=1}^N 
\left( 4\tau_1(n) + 2\tau_2(n) \right) = N\tau_0 = T,~~\mbox{and}
\nonumber \\
T^{CD}(\tau_0) &= 
N\tau_0\left[1+\frac{1}{\delta \tau_0}+ \frac{1}{N}\sum_{n=1}^N \kappa_n 
\right] = f_{\delta,J,N,T} T.
\end{align}
In our experiments, the factor $f_{\delta,J,N,T}$ ranged from 1.9 to 72.5. For a fair comparison,  we adjusted $\tau_0$ independently in 
AD and CD experiments to ensure that the physical implementation time durations of AD evolution matched that of CD evolution, i.e., $T^{AD} = T^{CD}$. 
The barplot of Fig. \ref{cd_ad_bar} (a) compares the experimental LLS signals of AD and CD for the same set of total durations $T$ as in the numerical simulations shown in Fig. \ref{fig:fidelity}.  Each bar is obtained by averaging five experiments corresponding to different iteration numbers $N \in \{10,15,20,25,30\}$.  
We find that CD performs better than AD at shorter times, as predicted by the numerical simulations in Fig. \ref{fig:fidelity}. At the larger total time $T=0.5$ s, CD appears to outperform AD, again in agreement with the numerical simulation. However, at intermediate times, AD seems to perform better.  

The barplot of Fig. \ref{cd_ad_bar} (b) compares the experimental LLS signals of AD and CD for the same set of the total number of iterations $N$, wherein each bar is obtained by averaging five different experiments corresponding to different total times $T \in \{0.01,0.05,0.1,0.3,0.5\}$. In  Fig. \ref{cd_ad_bar} (b), we see a similar trend as in  Fig. \ref{cd_ad_bar} (a), and CD performs better at low and high iteration numbers while AD performs better at intermediate iteration numbers. We find that the performances of AD and CD are complementary to each other.
The deviations from the simulations in Fig. \ref{fig:fidelity} could be due to the truncated WALTZ-16 cycles during CD segments and and other experimental imperfections such as RF inhomogeneity, spin relaxation, etc. which were not accounted for 
in numerical simulations.

Fig. \ref{cd_ad_bar} (c) shows the LLS signal as a function of the storage time. The LLS decay-constant is 27.3 s, which is over five times the longitudinal relaxation time-constant $T_1$ of 5.1 s for individual spins, confirming the long-livedness of the singlet state.



\section{Conclusions}
Adiabatic driving (AD) is important for numerous applications, from spectroscopy to upcoming quantum technologies. However, ensuring no transitions and maintaining adiabaticity often requires long driving durations. This can be a significant limitation for many applications, including today's noisy intermediate-scale quantum (NISQ) technologies. One promising way to overcome this limitation is to employ shortcuts to adiabaticity (STA) protocols. Counterdiabatic driving (CD) is a type of STA protocol that involves including an auxiliary component to the driving Hamiltonian to cancel the diabatic transitions by maintaining the moving frame Hamiltonian to be diagonal and inertial.  

In this work, after a brief review of AD, we outlined the theory of CD, and explained it using a simple two-qubit system. We compared the evolution of eigenvalues under the AD and CD evolutions. We observed the widening of energy gaps in the CD evolution, which explains how CD can avoid diabatic transitions even at faster driving.

Using a two-qubit NMR system, we experimentally studied AD and CD for preparing long-lived singlet states (LLS). LLS is recently being used for several spectroscopic and biomedical applications \cite{pileio2020long}. We described NMR pulse sequences to initialize, to prepare LLS via AD and CD evolutions, as well as to convert LLS into observable magnetization. We used numerical simulations to compare AD and CD pulse decompositions with ideal evolutions. Finally, we experimentally implemented these sequences on the two-qubit NMR system over a set of total time durations and a set of the total number of iterations. The results confirmed that CD achieves better LLS preparation with faster driving than AD. In general, the performances of AD and CD were found to be complementary. 

Already there are attempts to use deep learning \cite{PhysRevX.11.031070}, variational methods \cite{doi:10.1098/rsta.2021.0282}, and optimal control methods \cite{PhysRevLett.114.177206,_epait__2023} to optimize CD paths. We expect such developments in quantum control with advanced driving strategies to have important roles in the future.  

\begin{acknowledgments}
The authors acknowledge valuable discussions with Conan  Alexander.  
A.S. acknowledges fellowship from KVPY program of DST.
V.V. and T.S.M. acknowledge the funding  from DST/ICPS/QuST/2019/Q67. 
P.B. acknowledges support from the Prime Minister's Research Fellowship (PMRF) of the Government of India. We thank the National Mission on Interdisciplinary Cyber-Physical Systems for funding from the DST, Government of India, through the I-HUB Quantum Technology Foundation, IISER-Pune.  
\end{acknowledgments}

\section*{Data Availability Statement}
The data that support the findings of
this study is available from the
corresponding author upon reasonable
request.

\bibliographystyle{ieeetr}
\bibliography{citations}
\end{document}